\documentclass[12pt]{article}

\renewcommand{\title}[1]{\begin{center}\bf\Large #1\end{center}}

\renewcommand{\author}[1]{\begin{center}\large #1\end{center}}

\usepackage{latexsym,amsfonts}
\newcommand{\rr}{\mathbb{R}}

\begin{document}

\title{Zero Mode Problem of Liouville Field Theory}

\author{
 George Jorjadze${}^a$
\footnote{email: \tt jorj@rmi.acnet.ge}
 and Gerhard Weigt${}^b$
\footnote{email: \tt weigt@ifh.de} \\
{\small${}^a$Razmadze Mathematical Institute,}\\
  {\small M.Aleksidze 1, 380093, Tbilisi, Georgia}\\
{\small${}^b$DESY Zeuthen, Platanenallee 6,}\\
{\small D-15738 Zeuthen, Germany}}
\begin{abstract}
  We quantise canonical free-field zero modes $p$, $q$ on a half-plane
  $p>0$ both, for the Liouville field theory and its reduced Liouville
  particle dynamics. We describe the particle dynamics in detail,
  calculate one-point functions of particle vertex operators, deduce
  their zero mode realisation on the half-plane, and prove that the
  particle vertex operators act self-adjointly on a Hilbert space
  $L^2(\rr_+)$ on account of symmetries generated by the $S$-matrix.
  Similarly, self-adjointness of the corresponding vertex operator of
  Liouville field theory in the zero mode sector is obtained by
  applying the Liouville reflection amplitude, which is derived by the
  operator method.

\vspace{0.3cm}
$~~~~~~~~~~~~~~~\,$Liouville particle
 dynamics; Zero modes; Half-plane 
 quantisation;  


\end{abstract}

\baselineskip=20pt

\section{Introduction}
Although Liouville field theory was intensively studied in different 
areas of mathematics and physics, its quantum mechanical picture
remained incomplete with respect to some fundamental questions.  But the
rather simple form of the general solution \cite{L}, its free field
representations \cite{JPP}-\cite{OW} and several already obtained
exact quantum results \cite{BCT}-\cite{JW} are promising indications
for a complete description.  The unsolved problem we are
refering to is the quantisation of the canonical free-field zero modes
$p$, $q$ on a half-plane $p>0$ \cite{BCT, JW} which describe the
vacuum of regular periodic Liouville field theory.
This half-plane problem is a consequence
of the Liouville dynamics.

In this paper, we quantise these canonical zero modes, and demonstrate,
in particular, the self-adjoint action of vertex operators on a
Hilbert space $L^2(\rr_+)$ by taking into consideration symmetries
generated by an $S$-matrix. We have in view to complete our previous
work on periodic Liouville theory \cite{JW} in order to be able to
calculate correlation functions.  Moreover, solving this zero mode
problem would allow us, as well, to develop further the quantisation
of the $SL(2,\rr)/U(1)$ black hole model treated in ref.  \cite{FJW}.
In fact such investigations are fundamental for the quantisation of
gauged WZNW theories. Our calculations are based on an exact
canonical operator formalism in a Hilbert space. However, we do not
follow here our previous suggestion \cite{JW} to use a coherent state
formalism which considers translation and dilatation symmetry of the
half-plane.

We investigate the zero modes by using a canonical map
from a plane to a half-plane. Such a map naturally arises for particle
dynamics in an exponential potential by passing from the particle
phase-space coordinates to their asymptotic ($in$- or $out$-)
variables. The particle approach seems to be useful because Liouville
field theory can be reduced to a mechanical model, and much
is already known about Liouville particle dynamics
\cite{Jack}, \cite{BCGT}-\cite{F}.  But the quantisation on the
half-plane was not sufficiently discussed so far.  In particular, the
self-adjoint action of $q$-exponentials on $L^2(\rr_+)$ is not 
understood and the use of Liouville vertex operators 
for the calculation of Liouville correlation functions is therefore
not yet fully justified. The detailed treatment of Liouville
particle dynamics will prove to be very helpful to understand
the zero mode structure of Liouville field theory.

In Section 2 we summarise the free field representation of the
periodic Liouville theory and define the zero mode problem. Here we
describe the classical and quantum mechanical reduction of the
Liouville theory to particle dynamics, and deliver the canonical
transformation to the free-field zero modes on the half-plane. In
Section 3 we elaborate the quantum particle dynamics in the
exponential potential, use the M{\o}ller matrix to get the quantum
realisation of the relevant zero mode operators, calculate matrix
elements of the mechanical Liouville vertex operator and extract from
them the particle vertex operator in terms of the half-plane zero
modes.  Here a smooth continuation in the coupling has to be performed
of the matrix elements which have to be considered as generalised
functions. The self-adjoint action of vertex operators on the Hilbert
space $L^2(\rr_+)$ can be realised both, for the particle case and the
zero mode sector of field theory, by using symmetries of the vertices
under reflection.  The reflection amplitude of the Liouville field
theory is derived by the operator method. It is surprisingly identical to
that deduced in \cite{ZZ} by a symmetry of a 3-point correlation
function suggested in \cite{DO} on account of path-integral
considerations.  We summarise the results and conclude.  Some details
are given in Appendices.

\section{Zero modes of periodic Liouville theory}

To obtain a useful definition of the zero mode problem, we report
the canonical free-field parametrisation of the periodic Liouville
theory, reduce it to Liouville particle dynamics,
classically and quantum mechanically, and observe
that the particle phase-space coordinates are asymptotically 
related to the free-field zero modes on the half-plane by means of a canonical
transformation which is the classical analogue of the M{\o}ller matrix.

\subsection{Periodic Liouville field theory}

Solutions of the Liouville equation
\begin{equation}\label{Liouville-equation}
\varphi_{\tau\tau}(\tau,\sigma) 
-\varphi_{\sigma\sigma}(\tau,\sigma)
+\frac{4m^2}{\gamma}\,\, e^{2\gamma\varphi(\tau,\sigma)}=0
\end{equation}
have the useful free-field representation \cite{JPP},\cite{Jack}
\begin{equation}\label{Liouville-free}
e^{-\gamma\varphi(\tau,\sigma)}=e^{-\gamma\phi(\tau,\sigma)}
\left(1+m^2
\int\int dy\,d\bar y\,G(z,\bar z;\,y,\bar y)
\,e^{2\gamma\phi(y)}\,e^{2\gamma\phi(\bar y)}\right),
\end{equation}
where $\gamma$ and $m$ are positive constants, $z=\tau +\sigma$ and
$\bar z=\tau -\sigma$ light-cone coordinates,
$\phi(\tau,\sigma)=\phi(z)+\bar\phi(\bar z)$ is a chirally decomposed
free field, and $G(z,\bar z;\,y,\bar y)$ the Green's function of the
D'Alembert operator $\partial^2_{z\bar z}$. For periodic boundary
conditions $\varphi(\tau,\sigma+2\pi)=\varphi(\tau,\sigma)~$ the range
of integration in (\ref{Liouville-free}) is $(y ,\bar y)\in
[0,2\pi]\times[0,2\pi]$, the parametrising free field
$\phi(\tau,\sigma)$ is also periodic and it can be expanded in Fourier
modes
\begin{equation}\label{mode-exp}
\phi (\tau,\sigma)= q+
\frac{p}{2\pi}\tau +
\frac{i}{\sqrt {4\pi}}\sum_{n\neq 0}\frac{1}{n}\,\left(
{a_n}\, e^{-in(\tau +\sigma)}
+ {\bar a_n}\, e^{-in(\tau -\sigma)}\right ).
\end{equation}
Canonical Poisson brackets for the zero modes and oscillators
\begin{equation}\label{mode-PB}
\{ p,\, q\} =1,~~~~~\{a_m,\, a_n\}=im\delta_{m+n,0}=
\{\bar a_m,\, \bar a_n\}
\end{equation}
define relation (\ref{Liouville-free}) as a canonical transformation
between the Liouville and the free field.

For periodic boundaries 
the Green's function $G$ of (\ref{Liouville-free}) can be expressed by \cite{OW}
\begin{equation}\label{G}
G(z,\bar z;\,y,\bar y)= \theta_{\gamma p}\, (z-y)\theta_{\gamma
  p}\, (\bar z-\bar y),
\end{equation}
where  
\begin{equation}\label{theta_lambda}
\theta_{\gamma p} (z-y)=\frac{e^{\frac{\gamma p}{2}\, 
\epsilon(z-y)}}{2\,\sinh\frac{\gamma p}{2}}
\end{equation}
is the Green's function which inverts the operator $\partial_z$ on
functions $A(z)$ with the monodromy $A(z+2\pi)=e^{\gamma p}A(z)$, for
$p\neq 0$. $\epsilon (z)$ is the stair-step function. Its
differentiation gives the periodic $\delta$-function. The point $p=0$
is singular and will be excluded from (\ref{theta_lambda}). One can
show that $p>0$ (or $p<0$) covers the class of all regular periodic
Liouville fields, and in order to avoid double counting of the
solution (\ref{Liouville-free}) the zero modes are assumed to live on
the half-plane $p>0$ only. This dynamical restriction requires however
additional investigations for the quantisation \cite{JW}. The Hilbert
space of the non-zero modes $a_n$, $\bar a_n$ is the usual Fock space,
whereas the Hilbert space of the free-field zero modes is defined by
wave functions $\psi(p)\in L^2(\rr_+)$ .  Unfortunately, the standard
coordinate operator in the momentum representation $\hat q=i\hbar
\partial_p$ is not self-adjoint on $L^2(\rr_+)$, and the operators for
the exponentials $e^{\pm \gamma q}$ which appear in
(\ref{Liouville-free}) have to be suitably defined for applications.
This hermiticity problem is related to the missing translation
symmetry on the half-plane $p>0$ in `$p$-direction'. But we are going
to show that $S$-matrix transformations allow to `restore' this symmetry
by replacing $L^2(\rr_+) \rightarrow L^2(\rr^1)$, where
self-adjointness of $e^{\pm \gamma \hat q}$ is obvious.

\subsection{Reductions of Liouville theory to 
particle dynamics}

We can reduce the periodic Liouville theory to a mechanical Liouville
model by different methods. Since Liouville vacuum configurations
are described by a homogeneous field
$\partial_\sigma\varphi(\tau,\sigma)=0$, the belonging Liouville field
 is a time-dependent coordinate only $\varphi(\tau,\sigma)=Q(\tau)$ and
the Liouville equation (\ref{Liouville-equation}) reduces to the
mechanical Liouville model
\begin{equation}\label{m-Liouville}
\ddot{Q}(\tau) +\frac{4m^2}{\gamma}\, e^{2\gamma Q(\tau)}=0.
\end{equation}
The reduced Hamiltonian becomes
\begin{equation}\label{m-Hamiltonian}
H=\frac{1}{4\pi}\left(P^2+4\omega^2\, e^{2\gamma Q}\right),
\end{equation}
where $P=2\pi\dot{Q}$ is the canonical conjugated momentum,
and the parameter $\omega=2\pi m/\gamma$. 
The chiral energy-momentum tensor $T(z)=(\partial_z\varphi)^2-
\frac{1}{\gamma}\partial_{zz}^2\varphi$ is then a positive
constant $T(z)=c^2$, and its free-field representation
$T(z)=\phi\,'^2(z)-\frac{1}{\gamma}\phi\,''(z)$ leads to
$\phi\,'(z)=c$, which requires vanishing 
oscillator modes $a_n=0=\bar a_n$. As a consequence, the
Liouville solution (\ref{Liouville-free})-(\ref{theta_lambda}) reduces to
the general solution of the particle equation (\ref{m-Liouville})
\begin{equation}\label{g-solution}
e^{-\gamma Q(\tau)}=
e^{-\gamma \left(q+\frac{p\tau}{2\pi}\right)} +
\frac{\omega^2}{p^2}\,\,e^{\gamma \left(q+\frac{p\tau}{2\pi}\right)}.
\end{equation}

It is obvious that the mechanical Liouville model can be obtained also
vice versa from Liouville theory by Hamiltonian reduction with the
second class constraints $a_n=0=\bar a_n$, and that the reduced Hamiltonian
(\ref{m-Hamiltonian}) is canonically related by (\ref{g-solution}) 
to the free form
\begin{equation}\label{m-Hamiltonian1}
H=\frac{p^2}{4\pi}.
\end{equation}

It is worth mentioning that the above result can be obtained
directly from $SL(2,\rr)$ WZNW theory \cite{Dublin, JW} or its
homogeneous form, the $SL(2,\rr)$ particle model \cite{F}, by a
suitable Hamiltonian reduction.  
If we identify the gauge invariant part of the $SL(2,\rr)$ field,
the  $g_{12}\,(\tau)$ of \cite{JW}, with the particle vertex function 
$V(\tau)=e^{-\gamma
  Q(\tau)}$, from the $SL(2,\rr)$ particle equations
of motion
\begin{equation}\label{e}
\ddot g_{\alpha\beta}\,(\tau) =
\frac{\gamma^2}{\pi}\,H\, g_{\alpha\beta}\,(\tau)
\end{equation}
one reads off the 
gauge invariant equation 
\begin{equation}\label{v-eq}
\ddot V(\tau) =\frac{\gamma^2}{\pi}\,H\, V(\tau),
\end{equation}
which is equivalent to (\ref{m-Liouville}).
Its integration reproduces the general
solution (\ref{g-solution}) by using (\ref{m-Hamiltonian1}) in the form
\begin{equation}\label{p-H}
p=\sqrt{4\pi H}.
\end{equation}

It might be interesting to discuss the
 relic of the conformal symmetry of Liouville theory
\begin{equation}\label{confsymmetry}
\varphi(z,\bar z)\rightarrow
\varphi(f(z),\bar f(\bar z))+\frac{1}{2\gamma}
\log\, f\,'(z)\,\bar f\,'(\bar z).
\end{equation}
For $\sigma$-independent Liouville fields
the functions
$f(z)$ and $\bar f(\bar z)$ can  be linear  only, and with
the parametrisation $f(z)=az+b$, $\bar f(z)=a\bar z+b$
(\ref{confsymmetry}) reduces to 
\begin{equation}\label{symmetry}
Q(\tau)\rightarrow Q(\tau +b),~~~~~~~~~~~~~
Q(\tau)\rightarrow Q(a\tau)+\frac{\log a}{\gamma},~~~~~a>0. 
\end{equation}
These transformations are symmetries of the dynamical equations 
(\ref{m-Liouville}) respectively (\ref{v-eq}), and they
transform the zero modes $p,\,q$ of (\ref{g-solution}) as
\begin{equation}\label{symmetry-pq}
(\,p,\,q\,)\rightarrow (\,p,\,q+\frac{pb}{2\pi}\,),
~~~~~~~~~~~
(\,p,\,q\,)\rightarrow (\,a\,p,\, q+\frac{\log a}{\gamma}\,). 
\end{equation}
The time translations are obviously generated by the 
free Hamiltonian (\ref{m-Hamiltonian1}), whereas the remaining dilatation are
not canonical anymore, since they transform the action by
$S\rightarrow a\,S$ which is not a Noether symmetry.

In Appendix A, for amuse, we show that 
the mechanical Liouville model
has a nice relativistic free-particle
interpretation.

\subsection{Asymptotics of mechanical Liouville model}

The mechanical
Liouville model  becomes asymptotically a free theory. The general
solution (\ref{g-solution}) delivers
\begin{equation}\label{in-varibales}
\lim_{\tau \rightarrow -\infty}\,
\left[Q(\tau) - q(\tau)\right]=0,
~~~~~~~~~~\lim_{\tau \rightarrow -\infty}\,\left[ P(\tau)-p\right] =0,
\end{equation}
where $q(\tau)=q+\frac{p\tau}{2\pi}$ is the free-particle solution.
Therefore, $p$ and $q$ can be interpreted as the $in$-variables of the
Liouville particle dynamics. Similarly we introduce $out$-variables by
the behaviour of (\ref{g-solution}) as $\tau \rightarrow +\infty$.
The $in$- and $out$-variables are related by
\begin{equation}\label{out-varibales}
P_{out}=-p, ~~~~~~Q_{out}=-q+\frac{2}{\gamma}\log\,\frac{p}{\omega},
\end{equation}
which combines the reflection of $p,\,q$ with a canonical map.
 It is an important observation that the general solution (\ref{g-solution})
is invariant under the transformation (\ref{out-varibales}).
Quantum mechanically this symmetry will be
generated by the $S$-matrix of the particle theory. Note that in Liouville field
theory $in-$ and $out-$fields \cite{JPP} are related 
by the special M\"obius transformation
$A\rightarrow -(m^2\,A)^{-1},~~ 
\bar A\rightarrow -(m^2\,\bar A)^{-1}$, which keeps invariant 
 the general solution
\begin{equation}\label{Liouville-solution}
\varphi(\tau,\sigma) =\log \frac{A\,'(\tau +\sigma)\bar A\,'(\tau -\sigma)}
{[1+m^2 A(\tau +\sigma)\bar A(\tau -\sigma)]^2}\,.
\end{equation} 

Due to (\ref{g-solution}), the phase space coordinates $P,\,Q$ 
at $\tau=0$ and the asymptotic variables
$p,\,q$ are related by
\begin{eqnarray}\label{PQ-pq}
e^{-\gamma Q}=
e^{-\gamma q} +
\frac{\omega^2}{p^2}\,\,e^{\gamma q},~~~~~~~~~
P=-p\,\tanh \left(\gamma q+\log\frac{\omega}{p}\right).
\end{eqnarray}
This is a canonical and invertible map of the plane $(P,Q)$ onto the
half-plane $(p,q)$. Its inversion gives (\ref{p-H}) and provides $q$
as a function of $P,\,Q$. The transformation can be simplified for
the coordinate $\tilde q=q+\frac{1}{\gamma}\,\log \frac{\omega}{p}$
\begin{eqnarray}\label{PQ-pq-PQ}
&&e^{-\gamma Q}=\frac{2\omega}{p}\,\cosh \gamma\tilde q,
~~~~~~~~~~~~~~~~2\omega\sinh \gamma\tilde q =-P\,e^{-\gamma Q}; \nonumber \\
&&P=-p\,\tanh \gamma \tilde q,~~~~~~~~~~~~~~~~~~~~p=\sqrt{
P^2+4\omega^2\,e^{2\gamma Q}},
\end{eqnarray}
and summarised by the generating function $F(Q,\tilde q)=
\frac{2\omega}{\gamma}\,e^{\gamma Q}\sinh \gamma \tilde q~$ of ref. 
\cite{CFZ}.

Properly speaking, the half-plane problem becomes visible already on
the $(P,Q)$-plane, where the Hamiltonian (\ref{m-Hamiltonian})
continuously relates the positive $in$- and negative $out$- momenta.
This is illustrated in Fig.$\,$1. Each phase-space trajectory of the
full $(P,Q)$-plane is one-to-one mapped by (\ref{PQ-pq-PQ}) to a
corresponding line $p=const$ on the upper half-plane $(p,q)$ of
Fig.${\,}$2.  We observe a `non-degeneracy' on the phase-space
$(P,Q)$, which quantum mechanically will be described by the
non-degeneracy of the Hamiltonian spectrum.

\setlength{\unitlength}{.1mm}

\begin{picture}(1500,520)

\put(0,240){\bf\line(1,0){600}}
\put(280,80){\line(0,1){365}}

\put(840,240){\line(1,0){600}}
\put(1140,80){\line(0,1){370}}

\put(261,450){$P$}
\put(610,235){$Q$}

\put(1125,462){$p$}
\put(1455,240){$q$}

\put(60,320){\line(1,0){300}}
\put(60,300){\line(1,0){200}}
\put(60,270){\line(1,0){100}}

\put(60,210){\line(1,0){100}}
\put(60,180){\line(1,0){200}}
\put(60,160){\line(1,0){300}}

\put(200,240){\oval(450,160)[r]}
\put(150,240){\oval(400,120)[r]}
\put(100,240){\oval(250,60)[r]}

\put(885,280){\line(1,0){510}}
\put(885,340){\line(1,0){510}}
\put(885,400){\line(1,0){510}}

\put(300,310){$\rightarrow$}
\put(180,290){$\rightarrow$}
\put(120,260){$\rightarrow$}

\put(120,200){$\leftarrow$}
\put(180,170){$\leftarrow$}
\put(300,150){$\leftarrow$}

\put(990,270){$\rightarrow$}
\put(990,330){$\rightarrow$}
\put(990,390){$\rightarrow$}

\put(230,245){$^1$}
\put(355,245){$^2$}
\put(435,245){$^3$}

\put(855,290){$_1$}
\put(855,350){$_2$}
\put(855,410){$_3$}

\put(240,30){$\mbox{Fig.\,1}$}
\put(1100,30){$\mbox{Fig.\,2}$}
\end{picture}

Unfortunately, the map (\ref{PQ-pq}) is non-linear and its
quantum realisation becomes non-trivial. But there is a
limiting procedure which will simplify the quantum calculations. For
this purpose we introduce an additional free Hamiltonian $H_0=P^2/4\pi$ and
consider the Hamiltonian flows $U_{H_0}(\tau)$ and $U_H(\tau)$
generated by $H_0$ and $H$ respectively
\begin{equation}\label{U_Ht}
U_{H_0}(\tau)\, (P,\,Q)=(P,\,Q+\frac{P\,\tau}{2\pi}\,),~~~~~~~~
U_{H}(\tau)\, (P,\,Q)=(P(\tau),\,Q(\tau)\,).
\end{equation}
Here $P(\tau),\,Q(\tau)$ describe the solution of Hamiltonian equations
with the initial data $P,\,Q$. The composition of the two flows
$U(\tau)=U_{H}(\tau)\cdot U_{H_0}(-\tau)$ yields
\begin{equation}\label{U-t}
U(\tau)\, (P,\,Q)=(P(\tau),\,Q(\tau)-\frac{P(\,\tau)\tau}{2\pi}),
\end{equation}
and because of (\ref{in-varibales}) we obtain the canonical map of the
$(P,Q)$-plane onto the half-plane $(p,q)$ by the asymptotic relation
\begin{equation}\label{e^Ht}
 \lim_{\tau\to -\infty} U(\tau)\,(P,Q)=(p,q).
\end{equation}
Quantum mechanically this map is mediated by the M{\o}ller matrix.
 
Note that $U(\tau)$ defines a canonical transformation of the 
coordinates $P,\,Q$ for any $\tau$.
We also mention that (\ref{theta_lambda}) yields
$G(z,\bar z;\,y,\bar y)\rightarrow 0$ as $\tau
\rightarrow -\infty$, and the parametrising free field
(\ref{mode-exp}) is therefore an asymptotic $in$-field of the
Liouville theory. The Liouville zero mode sector is so consistently
given by the $in$-variables of the mechanical model.

\subsection{Reduction of quantum Liouville field theory}

We only sketch 
the reduction of quantum Liouville theory to quantum particle
dynamics.

The free field representation
(\ref{Liouville-free})-(\ref{theta_lambda}) was used in \cite{OW} for
an algebraic construction of the Liouville vertex operator
$\left(e^{\lambda\varphi(\tau,\sigma)}\right)_{op}$.  With the
definition of the Liouville operator field $\hat \varphi
(\tau,\sigma)= \partial_\lambda
\left(e^{\lambda\varphi(\tau,\sigma)}\right)_{op}|_{\lambda =0}$ it
was shown that the operator Liouville equation and the canonical
commutation relations remain valid, although the free-field
description of the Liouville operators is deformed as compared with
the classical analogue.  Both, the oscillator and zero mode
operators contribute jointly to these deformations and yield conformal
invariant, local and anomaly-free operator structures. These
deformations also guarantee, e.g., that the $4$-point function of the
$25$-dimensional non-critical string becomes consistently the critical
Shapiro-Virasoro amplitude in $26$-dimensional Minkowski space-time.

The mechanical Liouville model is obtained by Hamiltonian reduction of
the Liouville field theory with respect to the constraints
$a_n=0=\bar a_n$.  One can simulate such a reduction quantum mechanically
by calculating oscillator vacuum matrix elements of 
vertex operators
\begin{equation}\label{V(0,0)}
{\bf \hat V}_\lambda(\tau)=\langle\,{\bf 0} |
\left(e^{\lambda\varphi(\tau,\sigma)}\right)_{op}
|\, {\bf 0}\rangle,
\end{equation}
where the vacuum state $|\, {\bf 0}\rangle$ is annihilated by
the oscillators $\hat a_n$, $ \hat{\bar{a}}_n$ for $n>0$.

This matrix element is obviously an operator in the zero mode sector,
and one can prove its $\sigma$-independence. Then the Liouville
operator equation simply reduces to a particle operator equation. But
we should point out here that we have not disentangled the
oscillator and zero mode contributions by this reduction, since the
constraints $a_n=0=\bar a_n$ are of second class.  The result will be
nevertheless useful.

For the application it will be sufficient to treat the special Liouville vertex
operator for $\lambda =-\gamma$ only. Its explicit form 
is derived in Appendix B as
\begin{equation}\label{V(bb)}
{\bf \hat V}(\tau)
=e^{-\frac{\gamma\hat p}{4\pi}\tau}\,\, 
e^{-\gamma \hat q}\,\,e^{-\frac{\gamma\hat p}{4\pi}\tau}+
\omega_\alpha^2\,\,\frac{e^{\frac{\gamma\hat p}{4\pi}\tau}}{\hat p}\,\,
\frac{\Gamma\left(-i\frac{\gamma p}{2\pi}\right)}
{\Gamma\left(i\frac{\gamma p}{2\pi}\right)}\,\,
e^{\gamma \hat q}\,\,
\frac{\Gamma\left(i\frac{\gamma p}{2\pi}\right)}
{\Gamma\left(-i\frac{\gamma p}{2\pi}\right)}
\,\,\frac{e^{\frac{\gamma\hat p}{4\pi}\tau}}{\hat p},
\end{equation}
where 
\begin{equation}\label{omega_alpha1}
\omega_\alpha=\frac{2\pi m_\alpha\Gamma(1+2\alpha)}{\gamma}~~~~
\mbox{and}~~~~~m_\alpha^2 =\frac{\sin 2\pi\alpha}{2\pi\alpha}\,\,m^2\, . 
\end{equation}
Both, the deformed parameter $\omega_\alpha$ (\ref{omega_alpha}) and
the $\Gamma$-functions (\ref{integral}) are due to oscillator as well
as zero mode contributions, and (\ref{V(bb)}) has, of course, the
classical limit (\ref{g-solution}).

\section{Quantisation and self-adjoint vertex operators}

In this section we quantise the mechanical Liouville model, construct
the operators $\hat p$ and $e^{\pm\gamma\hat q}$ by means of the
M{\o}ller matrix, calculate matrix elements of particle vertex
operators and describe their realisation in terms of the zero mode
operators $\hat p$, $\hat q$.  Moreover, we shall investigate the
self-adjoint action of vertex operators on the Hilbert space
$L^2(\rr_+)$ using $S$-matrix properties, and derive, in particular,
the Liouville reflection amplitude by means of operator methods

\subsection{M{\o}ller matrix and operators of asymptotic zero modes}

For the quantisation of the canonical map of the $(P,Q)$-plane onto
the half-plane $(p,q)$ (\ref{e^Ht}) we introduce a M{\o}ller matrix
\cite{M}
\begin{equation}\label{Moler}
\hat U= \lim_{\tau\to -\infty} \hat U(\tau),
\end{equation}
which is given by the asymptotics of the unitary operator
\begin{equation}\label{U(t)}
\hat U(\tau)=
e^{\frac{i}{\hbar}\,\hat H\tau}e^{-\frac{i}{\hbar}\,\hat H_0\tau}.
\end{equation}
The operators corresponding to the two phase-spaces are then related by 
\begin{equation}\label{F(p,q)}
\hat p= \hat U  \hat P\hat U^+,~~~~~~
e^{\pm\gamma\hat q}= \hat U e^{\pm\gamma\hat Q} \hat U^+.
\end{equation}

To simplify the following calculations it is convenient to use the notation
\begin{equation}\label{alpha}
x=\gamma Q-\log\frac{\alpha}{m}~~~~\mbox{and}~~~~ 
\alpha =\frac{\hbar\gamma^2}{4\pi}.
\end{equation}
We write the Liouville and free Hamiltonian
operators in the $x$-representation
\begin{equation}\label{H,H_0}
\hat H=\hbar\alpha\left(-\partial^2_x+e^{2x}\right)~~~~~\mbox{and}~~~~~
\hat H_0=-\hbar\alpha\,\partial^2_x,
\end{equation}
which act on the Hilbert space $L^2(\rr)$.  The eigenstates of the
Hamiltonian $\hat H$ are the Kelvin functions \cite{Jack} discussed in
Appendix C.  In `bracket' notation these
eigenstates will be designed by $|\Psi_k\rangle$, and for the
eigenstates of $\hat H_0$ we use $|k\rangle$ ($-\infty<k<+\infty$).
Both eigenstates have the same normalisation $\langle k|k\,'\rangle
=\delta(k-k\,')=\langle \Psi_k|\Psi_{k\,'}\rangle$ and the same
eigenvalue $E=\hbar\alpha\,k^2\,\,(=p^2/4\pi)$.  But here we have to
emphasise that the spectrum of $\hat H_0$ is degenerated ($|k\rangle
\neq |{-k}\rangle$), whereas the spectrum of $\hat H$ is not
degenerated ($|\Psi_k\rangle=|\Psi_{-k}\rangle$); $|\Psi_k\rangle$ is
a complete set already for $k>0$ (see (\ref{normalisation})).  This
describes the quantum mechanical half-plane situation which we have
discussed in section 2.3 classically.

For the exponential operator ({\ref{U(t)}) we
  have so a `mixed' spectral decomposition
\begin{equation}\label{U(tau)}
\hat U(\tau) =\int_0^{+\infty}dk\int_{-\infty}^{+\infty}dk\,'\,
\, |\Psi_k\rangle \,U(k,k\,';\tau)\,\langle k\,'|,
\end{equation}
with $U(k,k\,';\tau) =e^{i\alpha(k^2-k\,'^2)\tau} \langle
\Psi_k|k\,'\rangle$.  In Appendix C is shown that the matrix 
element $\langle \Psi_k|k\,'\rangle$
is divergent and it has to be considered as a generalised function.
Its regularised form $\langle \Psi_k|k\,'\rangle_\varepsilon$ is given
 by (\ref{Fourier}).

With (\ref{Fourier}) and the well-known formulas of scattering theory 
\begin{equation}\label{scattaring}
 \lim_{t\to -\infty}\frac{e^{i(E-E')t}}{E-E'\pm i\varepsilon}
= -2\pi i\,\theta(\pm \epsilon)\delta(E-E'),
~~~~~~~E=\alpha k^2,
\end{equation}
the regularised kernel of (\ref{U(tau)}) has the limit
\begin{equation}\label{U(k,k')}
\lim_{\tau\to -\infty}U_\varepsilon(k,k\,';\tau)=
\,a_k\,\theta(k\,')\,\delta(k-k\,').
\end{equation}
According to (\ref{Psi_})-{(\ref{Psi-}), $a_k$ is the coefficient of
  the `left-moving' wave of the eigenfunction $\Psi_k(x)$ as $x\rightarrow
  -\infty$ with $|a_k|=1$; and $|\Psi_k^-\rangle
  =a_k\,|\Psi_k\rangle$ can be called therefore the $in$-eigenstate of the Hamiltonian
  $\hat H~$ \cite{S}. The M{\o}ller matrix then becomes
\begin{equation}\label{U}
\hat U=\int_0^{+\infty}dk\,|\Psi_k^-\rangle\, \langle k|,
\end{equation}
where the integration takes into consideration the non-degenerated
spectrum of the Hamiltonian $\hat H$ with $k>0$. The Hermitean
conjugated to the M{\o}ller matrix is obtained by similar calculations
\begin{equation}\label{U+}
\hat U^+=\lim_{\tau\to -\infty}
e^{\frac{i}{\hbar}\,\hat H_0\tau}e^{-\frac{i}{\hbar}\,\hat H\tau}=
\int_0^{+\infty}dk\,|k\rangle \,\langle \Psi_k^-|.
\end{equation}
It is important to note that the operator $\hat U(\tau)$ is unitary for 
finite $\tau$,
but in the limit $\tau\rightarrow -\infty$ only
a `one-side' unitarity condition remains
\begin{equation}\label{UU+}
\hat U \hat U^+=\hat I,~~~~~~~~~~\hat U^+\hat U=\int_0^{+\infty}dk
\,|k\rangle\,\langle k|.
\end{equation}
This non-unitarity is a consequence of the different spectra
of $\hat P$ and $\hat p$.

The transformation (\ref{F(p,q)}) formally preserves hermiticity, 
and with (\ref{U}),
(\ref{U+}) the spectral
decomposition of $\hat p$ on the half-line follows 
\begin{equation}\label{p}
\hat p = 
\hbar\gamma\int_0^{+\infty}dk\,|\Psi_k^-\rangle \,k\,\langle \Psi_k^-|.
\end{equation}
The momentum operator $\hat p$ is so self-adjoint,
$|\Psi_k^-\rangle $ is its eigenstate with eigenvalue
$p=\hbar\gamma k>0$, and  using again (\ref{U}), (\ref{U+}) one obtains
the useful relations
\begin{equation}\label{pU=UP}
\hat p\, \hat U = \hat U \,\hat P,~~~~~~~ 
\hat U^+\,\hat p = \hat P\,\hat U^+. 
\end{equation}
In the $x$-representation the operators $\hat P$ and 
$e^{\pm\gamma\hat Q}$ are 
\begin{equation}\label{P,eQ}
\hat P=-i\hbar\gamma\partial_x,
 ~~~~~~e^{-\gamma\hat Q}=\frac{m}{\alpha}\,e^{- x},
~~~~~~~e^{\gamma\hat Q}=\frac{\alpha}{m}\,e^{ x},
\end{equation}
and they satisfy the commutation relations 
$[\hat P,\,\hat e^{\pm\gamma\hat Q}]=\mp i\gamma\hbar e^{\pm\gamma\hat Q}.~$
Multiplying these relations on the left and 
right by $\hat U$ and $\hat U^+$ respectively, 
from (\ref{pU=UP}) we obtain
 a result consistent with canonical quantisation
\begin{equation}\label{[p,eq]}
[\hat p,\,\hat e^{\pm\gamma\hat q}]=\mp i\gamma\hbar e^{\pm\gamma\hat q}.
\end{equation}

We are also able to calculate the matrix elements $\langle
\Psi_k^-|\, e^{\pm\gamma\hat q}\,|\Psi_{k\,'}^-\rangle$.  Using
(\ref{F(p,q)}), (\ref{U})-(\ref{U+}), (\ref{P,eQ}) and the
orthonormality of $|\Psi_k^-\rangle$, they are given by
\begin{equation}\label{matrix-el}
\langle \Psi_k^-|\, e^{\pm\gamma\hat q}\,|\Psi_{k\,'}^-\rangle=
\left(\frac{\alpha}{m}\right)^{\pm 1}\langle k| e^{\pm\,\hat x}|k\,'\rangle.
\end{equation}
The matrix elements on the right hand side can be obtained
by means of the eigenstate $|k\rangle$ in
the $x$-representation,
$\sqrt{2\pi}\langle x|k \rangle= e^{ikx}$, as 
\begin{equation}\label{matrix-el1}
\langle k| e^{\pm\,\hat x}|k\,'\rangle=
\frac{1}{2\pi}\int_{-\infty}^{+\infty} dx \,e^{-i(k-k\,'\pm i)x}
=\delta(k-k\,'\pm i).
\end{equation}
The $\delta$-function with complex argument is defined on analytical 
test function in the standard way \cite{GSh}, and we will use
\begin{equation}\label{delta}
\int_0^{+\infty} dk\,'\delta(k-k\,'\pm i)
\psi (k\,')=\psi (k\pm i).
\end{equation}

Since $|\Psi^-_k\rangle$ is an eigenstate of $\hat p$, let us come
back to the initial zero mode variables and identify
$\sqrt{\hbar\gamma}\,\, |p \rangle= |\Psi^-_k\rangle$ for
$p=\hbar\gamma k$.  The eigenstates $|p\rangle$ are normalised
$\langle p|p\,'\rangle =\delta(p-p\,')$ and they define the
$p$-representation by wave functions $\psi(p)=\langle p|\Psi\rangle
\in L^2(\rr_+)$, which relates $\psi(p)$ with wave
functions of the $Q$-representation $\Psi(x)$ 
\begin{equation}\label{p-Q}
\psi(p) =\int_{-\infty}^{+\infty} dx\,\, 
\Psi_k^*(x)\,\Psi(x),~~~~~\mbox{for}~~~~~
 p=\hbar\gamma \,k.
\end{equation}
This transformation is unitary because the eigenstates $\Psi_k(x)$ are
complete. Then $\hat p\psi(p)=p\,\psi(p)$ and the operators
$e^{\pm\gamma\hat q}$ act due to (\ref{matrix-el}), (\ref{matrix-el1})
as
\begin{equation}\label{E}
e^{-\gamma\hat q}\,\psi(p)=\frac{m}{\alpha}\,\psi(p- i\hbar\gamma),~~~~~~~
e^{\gamma\hat q}\,\psi(p)=\frac{\alpha}{m}\,\psi(p+ i\hbar\gamma).
\end{equation}
These relations define the action of the zero mode operators in the
Hilbert space $L^2(\rr_+)$. But we have still to discuss the
hermiticity properties of the exponential $q$-operators, which will
be done in detail in section 3.3.

Note that any  operator of $F(p)$ given by 
a spectral decomposition like (\ref{p}) 
\begin{equation}\label{F(p)}
F(\hat p)=
\int_0^{+\infty}dp\,|p\rangle \,F(p)\,\langle p|
\end{equation}
satisfies relations similar to (\ref{pU=UP}), 
$F(\hat p)\,\hat U= \hat U F(\hat P)\,$ or
$\hat U^+ F(\hat p)=F(\hat P)\,\hat U^+$, and as a consequence
the multiplication rule holds
\begin{equation}\label{F(p)e^q}
F(\hat p)\,e^{\pm\gamma\hat q}=
e^{\pm\gamma\hat q}\,F(\hat p \mp i\hbar\gamma).
\end{equation}
This relation was used for the algebraic construction of
Liouville vertex operators in \cite{BCT,OW}.

\subsection{Vertex operators and their matrix elements}

The Heisenberg equations derived from
the Hamilton operator $\hat H=\hbar\alpha (-\partial_{xx}^2 +e^{2x})$
provide the operator equation
$\ddot{\hat x}(\tau) +4\alpha^2 e^{2\hat x(\tau)}=0$,
which has similarly as in
Liouville field theory the classical form (\ref{m-Liouville}).
However the particle vertex operator 
\begin{equation}\label{V(tau)}
\hat V(\tau)=e^{-\gamma\hat Q(\tau)}= 
e^{\frac{i}{\hbar}\,\hat H\tau}\,e^{-\gamma \hat Q}\,e^{-\frac{i}{\hbar}
\,\hat H\tau}
\end{equation}
satisfies a deformed equation
\begin{equation}\label{V(tau)1}
\frac{1}{2}\left(\ddot{\hat V}(\tau )\hat V(\tau )+
\hat V(\tau )\ddot{\hat V}(\tau )\right)-
\dot{\hat V}(\tau )^2+2\alpha^2\hat V(\tau )^2=4m^2,
\end{equation}
as compared with the corresponding classical one 
$\ddot V(\tau)V(\tau)-\dot V(\tau)^2=4m^2$. One observes
the ordering prescription for the term $\ddot V(\tau)V(\tau)$ and
a quantum correction proportional to $\alpha^2~$. Even 
the quantum version of the gauge invariant linear equation (\ref{v-eq}) has
a quantum deformation  
\begin{equation}\label{V(tau)2}
\ddot{\hat V}(\tau )=\frac{\gamma^2}{2\pi}\,(\,\hat H\, \hat V(\tau )+
\hat V(\tau )\,\hat H\,)+\alpha^2\, \hat V(\tau ).
\end{equation}
This parallels again Liouville theory, where 
the (anti-) chiral operators satisfy  
 quantum mechanically deformed Schr\"odinger
equations \cite{Neveu, OW}.

We are able to calculate matrix elements of vertex
operators of the mechanical model (Please, note that we also use
the notation $V$ for $V_{-1}$$\,$!) 
\begin{equation}\label{V(k,k)}
V_{2b}(k,k\,';\tau)=\langle\,\Psi_k^- |e^{{2b}\gamma\hat Q(\tau)}
|\,\Psi_{k\,'}^-\,\rangle.
\end{equation}
By (\ref{alpha})
and (\ref{d_k}), Eq. (\ref{V(k,k)}) becomes
\begin{equation}\label{e^Q(tau)1}
V_{2b}(k,k\,';\tau)=\left(\frac{\alpha}{m}\right)^{2b}\,d_k^*\,d_{k\,'}
\,\,e^{i\alpha(k^2-k\,'^2)\tau}\,\,\int_{-\infty}^{\infty}dx\,
K_{ik}(e^x)\,e^{2b x}\,K_{ik\,'}(e^x).
\end{equation}
But the integral is well defined for ${b} >0$ only, and it diverges for
${b} \leq 0$.  Eq. (\ref{e^Q(tau)1}) defines so a
generalised function which is a kernel of the vertex operator, and in
order to calculate it we need its smooth continuation from positive to
negative $b$.  Substituting (\ref{Kelvin}) into (\ref{e^Q(tau)1}) and
integrating over $u=e^x$, the integral splits into a product of
two integrals of the type (\ref{A(k)}) and gives with the notation
\begin{equation}\label{kappa}
\kappa= \frac{k+k\,'}{2}\,\,,~~~~~~~\rho=\frac{k-k\,'}{2}\,\,,
\end{equation}
for $b>0$  the result \cite{BCGT}
\begin{equation}\label{V(k,k)1}
V_{2b}=\left(\frac{\alpha}{m}\right)^{2b}\,\,e^{4i\alpha\rho\kappa\tau}\,4^{b-i\rho}\,
\frac{\Gamma\left(b+i\kappa\right)
\Gamma\left(b-i\kappa\right)}
{\Gamma\left(i\rho+i\kappa\right)
\Gamma\left(i\rho-i\kappa\right)}\,\frac
{\Gamma\left(b+i\rho\right)
\Gamma\left(b-i\rho\right)}{4\pi\,\Gamma(2b)}\,.
\end{equation}
In order to obtain the vertex function for negative $b$ one needs a
smooth continuation of (\ref{V(k,k)1}) as a generalised function,
which was not done before.
Since $\kappa>|\rho|$ ambiguities can arise at $b=-n$,
$n=0,1,2,...\,$. Near these points if $b=-n+\epsilon$ (\ref{V(k,k)1})
behaves like
\begin{equation}\label{delta+}
\frac{1}{\pi}\,\frac{\epsilon}{(k-k\,')^2+\epsilon^2}.
\end{equation}
We have shown in Appendix D that this generalised function has for
holomorphic test functions the following smooth continuation from
positive to negative values of $\epsilon$
\begin{equation}\label{delta-}
\frac{1}{\pi}\,\frac{\epsilon}{(k-k\,')^2+\epsilon^2}+
\delta(k-k\,'+ i\epsilon)+\delta(k-k\,'- i\epsilon).
\end{equation}

The continuation of (\ref{V(k,k)1}) to negative values of $b$ so
creates a pair of $\delta$-functions with complex arguments each time
$b$ passes a negative integer value, and for $b=-|b|$ results  
\begin{eqnarray}\label{V(k,k)2}
V_{-2|b|}=V_{-2|b|}^{+}\,(\kappa,\rho;\tau)
+~~~~~~~~~~~~~~~~~~~~~~~~~~~~~~~~~\nonumber\\
\left(\frac{m}{\alpha}\right)^{2|b|}\,\sum_{l=0}^{[|b|]}\, C_{2|b|}^l 
e^{-4\alpha(|b|-l)\kappa\tau}\,\,
\frac{\Gamma(-|b|+i\kappa)\,\Gamma(-|b|-i\kappa)}
{4^l\,\,\Gamma(-|b|+l+i\kappa)\,\Gamma(-|b|+l-i\kappa)}
\,\delta[2\rho-2i(|b|-l)] \nonumber\\
+\, C_{2|b|}^l\,e^{4\alpha(|b|-l)\kappa\tau}\,\,
\frac{\Gamma(-|b|+i\kappa)\,\Gamma(-|b|-i\kappa)}
{4^{2|b|-l}\,\,\Gamma(|b|-l+i\kappa)\,\Gamma(|b|-l-i\kappa)}
\,\delta[2\rho+2i(|b|-l)].
\end{eqnarray}
Here $V_{-2|b|}^{+}\,(\kappa,\rho;\tau)$ is defined by the r.h.s of
(\ref{V(k,k)1}) where $b$ is replaced by $-|b|$, $ [|b|]$ is
the integer part of $|b|$, and
\begin{equation}\label{C_b}
C_{2|b|}^l=\prod_{j=0}^{l-1}\frac{2|b|-j}{j+1}.
\end{equation}
In particular, the function $V_{-2|b|}^{+}$ vanishes at half-integer
$|b|$ due to the pole of $\Gamma(-2|b|)$, and it creates at integer
$|b|$ the terms proportional to $\delta(\rho)$, so that for $2|b|=n$
one has
\begin{equation}\label{V(k,k)3}
V_{-n}(\kappa,\rho;\tau)=\left(\frac{m}{\alpha}\right)^{n}\sum_{l=0}^nC_n^l\,
e^{-2(n-2l)\kappa\tau}\,\,\prod_{j=0}^{l-1} \frac{1}{4\kappa^2+(n-2j)^2}
\,\,\delta[2\rho-i(n-2l)],
\end{equation}
where the $C_n^l$ are now binomial coefficients. 

Let us consider the matrix element of the vertex operator $\hat V(\tau)$
which corresponds to the case $n=1$ of (\ref{V(k,k)3})
\begin{equation}\label{V(k,k,tau)}
V(k,k\,';\tau)=\frac{m}{\alpha}\,\left(
e^{-\alpha(k+k\,')\tau}\,\,\delta(k-k\,'-i)+
\frac{e^{\alpha(k+k\,')\tau}}{4\,k\,k\,'}\,\,\delta(k-k\,'+i)
\right). 
\end{equation}
From this expression and (\ref{E}) one can easily read off the
structure of the particle vertex operator as
\begin{equation}\label{e^Q=e^q}
\hat V(\tau)=
e^{-\frac{\gamma\hat p}{4\pi}\tau}\,\, 
e^{-\gamma \hat q}\,\,e^{-\frac{\gamma\hat p}{4\pi}\tau}+
\omega^2\,\frac{e^{\frac{\gamma\hat p}{4\pi}\tau}}{\hat p}\,\,
e^{\gamma \hat q}\,\,\frac{e^{\frac{\gamma\hat p}{4\pi}\tau}}{\hat p}.
\end{equation}
We observe here an ordering similar to the one of the reduced quantum
Liouville operator (\ref{V(bb)}), and by (\ref{F(p)e^q}) one can
easily check that the operator (\ref{e^Q=e^q}), indeed, satisfies both
the operator equations (\ref{V(tau)1}) and (\ref{V(tau)2}).

One can show that the vertex operator $\hat V_{-n}$ is the $n$-th
power of (\ref{e^Q=e^q}), however, for positive or non-integer
negative $2b$ the operator $\hat V_{2b}$ read off from (\ref{V(k,k)1})
or the first term of (\ref{V(k,k)2}) respectively is given (using
Fourier transformations) by an infinite series of $q$-exponentials, much as
in quantum Liouville field theory \cite{OW}.

\subsection{Self-adjointness of half-plane operators}

We have to ask whether (\ref{e^Q=e^q}), and (\ref{V(bb)}), are
physical operators with Hermitean action on the Hilbert space
$L^2(\rr_+)$.  Since asymptotically for $\tau\rightarrow\pm\infty$
only one term survives one is tempted to demand hermiticity for each
term of (\ref{e^Q=e^q}) separately.  But a proof of hermiticity for
$e^{\gamma \hat q}$ would require very special boundary conditions on
holomorphic functions $\psi(p)$ at $\mbox{Re}\,\, p=0$. This
mathematically by itself interesting but still unsolved problem
will be further discussed in Appendix E.

However, the vertex operator (\ref{e^Q=e^q}) is expected to become
self-adjoint as a whole 
on account of its symmetry under
transformations given by the $S$-matrix \cite{T}
\begin{equation}\label{S}
\hat S = \hat{\cal{P}}S(p).
\end{equation}
For the particle model $S(p)$ is the multiplicative reflection amplitude (\ref{s_k})
with $p=\hbar\gamma k$ and
$\hat{\cal{P}}$ the parity operator $\hat{\cal{P}}\psi(p)=\psi(-p)$.
It is easy to see that (\ref{S}) replaces the $in$-coming first term of 
(\ref{e^Q=e^q}) by the $out$-going second one, and vice versa,
so that the particle vertex operator remains invariant.
That means $\hat S \hat V (\tau)\hat S^{-1}$ is
identical to (\ref{e^Q=e^q}). This also holds in general as one can see 
from (\ref{V(k,k)1}) and (\ref{V(k,k)2}). 
Note that the $S$-matrix
is just the quantum version of the symmetry transformation
(\ref{out-varibales}).
$\hat S$ also maps the Hilbert space of the $in$-fields
$L^2(\rr_+)$ onto $L^2(\rr_-)$ for the $out$-fields, which for the wave 
functions $\psi(p)\in L^2(\rr_+)$, $\tilde\psi(p)\in L^2(\rr_-)$
is given by $\tilde\psi(-p)=S(p)\psi(p)$. The last relation is defined by 
(\ref{p-Q}) using (\ref{d_k}), (\ref{s_k}) and $\Psi_{-k}^*(x)=
d_{-k}^*\,K_{-ik}=S_{k}\,\Psi_{k}^*(x)$.

 Due to these properties one can extend the definition
of the matrix element of the vertex operator from the Hilbert space
of the half-plane
$L^2(\rr_+)$ to that of the plane $L^2(\rr)$
\begin{equation}\label{R+R}
\int_0^\infty dp\,\,\psi_+^*(p)\hat V(\tau)\psi_+(p)=\frac{1}{2}
\int_{-\infty}^\infty dp\,\,\Psi^*(p)\hat V(\tau)\Psi(p).
\end{equation}
Here $\Psi(p)=\psi(p)$ for $p>0$,
$\Psi(p)=\tilde\psi(p)=S(-P)\psi(-p)$ for $p<0$, and $\Psi(p)\in
L^2(R)$ satisfies $\hat S\Psi(p)=\Psi(p)$.  Self-adjointness of the
operator (\ref{e^Q=e^q}) obviously holds on $L^2(\rr)$ for holomorphic
wave functions $\Psi(p)$. As a consequence, self-adjointness of $\hat
V(\tau)$ on $L^2(\rr_+)$ requires for $\psi(p)$ a holomorphic
extension to the negative half-line so that
$\psi(-p)=\tilde\psi(-p)=S(p)\psi(p).$ Such functions are given by
$\psi(p)=d\,^*(p) f(p)$, where $d(p)$ is defined by (\ref{d_k}) with
$p=\hbar\gamma k$ and $f(p)$ is an even holomorphic function
$f(-p)=f(p)$. The integral (\ref{R+R}) is well-defined at the singular point
of $\hat V(\tau)$ $p=0$, since $\psi(0)=0$ on account of $d(0)=0$.

We can finally ask the question whether the Liouville vertex operator 
in the vacuum sector $\hat {\bf  V}(\tau)\,\,$ (\ref{V(bb)}) 
is invariant under the $S$-matrix transformation $\hat S_L=\hat{\cal P}S_L(p)$
 which exchanges the $in$-coming with the deformed $out$-going zero mode part.
We have solved the equation 
\begin{equation}\label{S0}
\hat S_L^{-1}\hat {\bf  V}(\tau)S_L=\hat {\bf  V}(\tau)
\end{equation}
and found exactly the deformed reflection amplitude of \cite{ZZ}
\begin{equation}\label{S1}
S_L(p) =-\left(\frac{4\pi m\alpha^2}{\Gamma(1+2\alpha)\,\sin 2\pi\alpha }
\right)^{\frac{2ip}{\hbar\gamma}}
\,\frac{\Gamma\left(i\frac{\gamma p}{4\pi\alpha}\right)}
{\Gamma\left(-i\frac{\gamma p}{4\pi\alpha}\right)}\,\
\frac{\Gamma\left(i\frac{\gamma p}{2\pi}\right)}
{\Gamma\left(-i\frac{\gamma p}{2\pi}\right)}\,.
\end{equation}
This result motivates one to look for a deeper understanding of the
path-integral based conjectures of 3-point functions by means of the
operator approach \cite{T, Th}. But for a proof of self-adjointness
the knowledge of the full Liouville $S$-matrix is required. The
operator equation for the Liouville $S$-matrix and its solution will
be discussed elsewhere.

\section{Summary and conclusions}

We investigated in detail Liouville particle dynamics which describes
Liouville field theory in the zero mode sector. Half-plane zero modes
have been quantised and proved to be self-adjoint on account of hidden
symmetries of vertex operators under $S$-matrix transformation. In
order to obtain matrix elements of particle vertex operators, a method
was developed to continue generalised functions smoothly in a coupling
parameter.

We learned from particle dynamics to treat quantum Liouville field
theory in the zero mode sector, calculated as an interesting result
the reflection amplitude by the operator method, and used it for the proof
of self-adjointness of corresponding vertex operators.

The analytical properties discovered for the particle matrix elements
are expected to be useful to relate results which have been derived
alternatively by the operator and path-integral approaches, so testing
the conjectured Liouville 3-point functions.

\vspace{0.3cm}
\noindent
{\bf {\Large Acknowledgements}}

\vspace{0.1cm}

\noindent
We thank T. Curtright, D. Fairlie, C. Thorn and C. Zachos for
bringing their work about Liouville particle dynamics to our
attention, and C. Ford, M. Reuter, W. R\"uhl, G. Savvidy, R. Seiler,
and J. Teschner for discussions.  G.J. is grateful to DESY Zeuthen for
hospitality.  His research was supported by grants from the DFG,
INTAS, RFBR and GAS.

\setcounter{equation}{0}
\def\theequation{A.\arabic{equation}}

\begin{appendix}

\section {Relation with a relativistic free-particle}

Here we give a relativistic  
free-particle interpretation of the Liouville particle model. 
For this purpose we introduce a 2-dimensional Minkowskian manifold with 
coordinates $X^0=T,\,X^1=X$ 
and the conformal metric tensor
\begin{eqnarray}\label{metric}
g_{\mu\nu}(T,X)=e^{2X}\left( \begin{array}{cr}   
  1&0\\0&-1 \end{array}\right).
\end{eqnarray}
 In this space-time the Lagrangian of a relativistic particle 
\begin{equation}\label{Lagrangian_R}
L=-m_0\,e^{X}\,\sqrt{\dot T^2-\dot X^2}
\end{equation}
 leads to the mass-shell condition
$P_T^2-P_X^2=m_0^2\,e^{2X}$. In the gauge $T=P_T\,\tau$ one obtains
a Liouville particle action 
\begin{equation}\label{action1}
S=\int \left(P_X\,dX-\frac{1}{2}(P_X^2+m_0^2e^{2X})d\tau\right). 
\end{equation}
The scalar curvature calculated 
for the metric (\ref{metric}) vanishes and one can pass to flat
coordinates $Y_\pm=
\pm e^{X\pm T}$, for which  (\ref{Lagrangian_R}) simply 
becomes the relativistic free
Lagrangian
\begin{equation}\label{L}
L=-m_0\sqrt{\dot Y_+\,\dot Y_-},
\end{equation}
 and it defines
straight-line trajectories
\begin{equation}\label{trajectory}
P_-Y_+-P_+Y_-=M. 
\end{equation}
Here $P_\pm$ are light-cone components of the relativistic momentum 
($P_+P_-=m_0^2$) and $M$ is the boost. Rewriting these trajectories
in $T, \,X$ coordinates and taking into account the gauge 
fixing condition $T=P_T\,\tau$,
one gets
\begin{equation}\label{g-solution1}
e^{-X(\tau)}=
\frac{P_+}{M}\,\,e^{-P_T\tau} +
\frac{P_-}{M}\,\,e^{P_T\tau}.
\end{equation}
This result reproduces the general solution (\ref{g-solution}),
and we have got a map of the solutions (\ref{g-solution}) to 
trajectories  of a free relativistic particle (\ref{trajectory}).
But since $Y_+>0$ and $Y_-<0$ these trajectories  
cover only one `quarter' of the space-time $Y^\mu$.
Note that the time translation symmetry of (\ref{symmetry}) corresponds
to Lorentz transformations of $Y^\mu$ coordinates.

\setcounter{equation}{0}
\def\theequation{B.\arabic{equation}}

\section {Reduced vertex operator}

We take the operator form of the Liouville vertex function
(\ref{Liouville-free}) from \cite{OW} (adapting the notation),
calculate the normal-ordered oscillator matrix element 
(\ref{V(0,0)}) for $\lambda=-\gamma$ 
\begin{equation}\label{Liouville-free1}
{\bf \hat V}(\tau)=:e^{-\gamma\left(\hat q+\frac{\hat p\tau}{2\pi}\right)}
\left(1+m_\alpha^2
\int_0^{2\pi}\int_0^{2\pi} dy\,d\bar y\,\,G_\alpha(z,\bar z;\,y,\bar y)
\,e^{2\gamma\left(\hat q+\frac{\hat p(y+\bar y)}{4\pi}\right)}\right):\,,
\end{equation}
and  use the ordering prescription 
\begin{equation}\label{ordering}
:e^{2aq}\,A(p):= e^{a\hat q}\,A(\hat p)e^{a\hat q}.
\end{equation}
The index $\alpha$ symbolises the quantum deformations, where
\begin{equation}\label{def-m}
m_\alpha^2 =\frac{\sin \pi\alpha}{\pi\alpha}\,\,m^2\,,~~~~\mbox{with}~~~~ 
\alpha =\frac{\hbar\gamma^2}{4\pi}.
\end{equation}
$ G_\alpha(z,\bar z;\,y,\bar y)$ is the Green's
function (\ref{G}),(\ref{theta_lambda}) deformed as
\begin{equation}\label{sinh}
\sinh^2\frac{\gamma p}{2}\rightarrow \sinh^2
\left(\frac{\gamma p}{2}\right)
+\sin^2\left(\frac{\hbar\gamma^2}{4}\right),
\end{equation}
and multiplied with the short distance factor
$f_\alpha(z-y)\,f_\alpha(\bar z-\bar y)$ given by
\begin{equation}\label{sin}
f_\alpha(x)=\left(4\sin^2\frac{x}{2}\right)^\alpha.
\end{equation}
We should mention here that the short distance factors are a consequence
of conformal invariance and the deformations (\ref{def-m}) and
(\ref{sinh}) are due to locality.
With the integral \cite{GR}
\begin{equation}\label{integral}
\int_0^{2\pi}\,dx\, \left(\sin \frac{x}{2}\right)^{2\alpha}
\,e^{\frac{\gamma px}{2\pi}}=
\,\frac{2^{2-2\alpha}\,\pi\, \Gamma(1+2\alpha)\,\,
e^{\frac{\gamma p}{2}}}{\Gamma
(1+\alpha+i\frac{\gamma p}{2\pi})\,\,
\Gamma(1+\alpha -i\frac{\gamma p}{2\pi})},
\end{equation}
and the identity
\begin{eqnarray}\label{Gamma.Gamma}
\Gamma\left(1-\alpha-i\frac{\gamma p}{2\pi}\right)
\Gamma\left(1-\alpha+i\frac{\gamma p}{2\pi}\right)
\Gamma\left(1+\alpha-i\frac{\gamma p}{2\pi}\right)
\Gamma\left(1+\alpha+i\frac{\gamma p}{2\pi}\right)
\\ \nonumber
=\frac{(\pi \alpha)^2+(\frac{\gamma p}{2})^2}
{\sinh^2 \frac{\gamma p}{2} +\sin^2\pi \alpha},
\end{eqnarray}
we arrive at (\ref{V(bb)}),
where 
\begin{equation}\label{omega_alpha}
\omega_\alpha=\frac{2\pi m_\alpha\Gamma(1+2\alpha)}{\gamma}
\end{equation}
contains all $p$-independent contributions, and it has 
the classical limit $\frac{2\pi m}{\gamma}$ of (\ref{g-solution}).

\setcounter{equation}{0}
\def\theequation{C.\arabic{equation}}

\section {Eigenstates of mechanical Liouville Hamiltonian}

The solutions of the stationary Schr\"{o}dinger equation 
\begin{equation}\label{H-Psi}
-\Psi_k\,''(x)+ e^{2x}\,\Psi_k(x)=k^2\,\Psi_k(x),
\end{equation}
are given by Kelvin (modified Bessel) functions 
 $\Psi_{k}(x)=c_k\,
K_{ik}\left(e^{x}\right)$ \cite{Jack} (see also Eqs. (2.16)-(2.22)
of \cite{BCGT}),
where $c_k$ is a normalisation coefficient. 
The Kelvin functions are real $K_{ik}^*(u)=K_{-ik}(u)=K_{ik}(u)$
and have the useful integral representation \cite{GR}
\begin{equation}\label{Kelvin}
K_{ik}(u)=\frac{1}{2}\,\int_{-\infty}^{+\infty}dy
\,e^{-u\,\cosh y}\,\,e^{iky}.
\end{equation} 
The eigenstates $\Psi_{k}(x)$  vanish rapidly for 
$x\rightarrow\infty$ and oscillate as 
$x\rightarrow -\infty$. The spectrum $E=\hbar\alpha k^2$ is non-degenerated, 
but the zero energy point $k=0$ has to be excluded from the 
spectrum since the corresponding wave function diverges as 
$x\rightarrow -\infty$, and for the eigenstates $\Psi_{k}(x)$
we take $k>0$ only.
 
The normalisation and completeness conditions 
\begin{equation}\label{normalisation}
\int_{-\infty}^{+\infty} dx\,\, \Psi_{k}(x)\Psi_{k\,'}(x) =\delta(k-k\,'),
~~~~~\int_{0}^{+\infty} dk\,\, \Psi_{k}(x)\Psi_{k}(x') =\delta(x-x\,')
\end{equation}
are valid \cite{KL} if
\begin{equation}\label{c_k}
c_k=\frac{1}{\pi}\sqrt{2\,k\,\sinh \pi k}.
\end{equation}
It is known from scattering theory that
the solutions of the stationary Schr\"odinger equation give
asymptotically complete information about a scattering process
\cite{S}.  The asymptotics of $K_\nu(u)$ for $u\rightarrow 0$
\cite{GR}
\begin{equation}\label{K_nu}
K_\nu(u)\rightarrow -\frac{\pi}{2}
\left[\frac{2^{-\nu}\,u^\nu}{\sin\pi\nu\,\,\Gamma(1+\nu)}
-\frac{2^{\nu}\,u^{-\nu}}{\sin\pi\nu\,\,\Gamma(1-\nu)}\right]
\end{equation}
delivers for the wave functions
\begin{equation}\label{Psi_}
\Psi_k(x)\rightarrow a_k^*\,\frac{e^{ikx}}{\sqrt{2\pi}}+
a_k\,\frac{e^{-ikx}}{\sqrt{2\pi}},~~~~\mbox{as}~~~x\rightarrow -\infty.
\end{equation}
The coefficient of the $out$-going wave
\begin{equation}\label{a_k}
a_k=-i\frac{2^{ik}}{\Gamma(1-ik)}\,\sqrt{\frac{\pi k}{\sinh\pi k}}
\end{equation}
is a phase due to the relation
\begin{equation}\label{GG}
\Gamma(1+ik)\Gamma(1-ik)=
\frac{\pi k}{\sinh\pi k}.
\end{equation}
The eigenstate $\Psi^-_k(x)=a_k\,\Psi_k(x)$ has therefore the asymptotics
\begin{equation}\label{Psi-}
\Psi_k^-(x)\rightarrow \frac{e^{ikx}}{\sqrt{2\pi}}+
S_k\,\frac{e^{-ikx}}{\sqrt{2\pi}},~~~~\mbox{as}~~~x\rightarrow -\infty.
\end{equation}
They satisfy obviously the normalisation and completeness conditions
(\ref{normalisation}), and  are given by
\begin{equation}\label{d_k}
\Psi^-_k(x)=d_k \,K_{ik}\left(e^{x}\right),~~~~~\mbox{with}~~~~
d_k=\sqrt{\frac{2}{\pi}}\,\frac{2^{ik}}{\Gamma(-ik)}.
\end{equation}
With (\ref{GG}) one easily finds for the reflection amplitude \cite{BCGT}
\begin{equation}\label{s_k}
S_k=2^{2ik}\,\frac{\Gamma(ik)}{\Gamma(-ik)}=\frac{d_k}{d^*_k}.
\end{equation}

The momentum representation of the eigenstates is 
defined by the Fourier transform of the Kelvin function
\begin{equation}\label{Psi_k}
\langle \Psi_k\,|\,k\,'\rangle
=\frac{1}{\sqrt{2\pi}}\int_{-\infty}^{+\infty} dx\, e^{ik\,'x}\, \Psi_k(x).
\end{equation}
This integral diverges 
and it has to be considered as a generalised function. 
We regularise (\ref{Psi_k})
by a factor $e^{\epsilon x}$, and using  (\ref{Kelvin}) as well as 
 the integral \cite{GR}
\begin{equation}\label{A(k)}
\int_{-\infty}^{+\infty}dy \frac{e^{iky}}{\cosh^\mu u}=
\frac{2^{\mu-1}}{\Gamma(\mu)}\, \,\Gamma\left(\frac{\mu+ik}{2}\right)
\Gamma\left(\frac{\mu-ik}{2}\right)~~~~~\mbox{for}~~~~~\mbox{Re}\,\mu>0,
\end{equation}
we get  the regularised momentum
representation of the eigenstate $\Psi_k(x)$ \cite{CFZ}
\begin{equation}\label{Fourier}
\langle \Psi_k\,|\,k\,'\rangle_\epsilon= 
c_k\,\frac{2^{\epsilon +ik\,'}}{4\sqrt{2\pi}}
\,\Gamma\left(\frac{\epsilon +ik\,'-ik}{2}\right)
\Gamma\left(\frac{\epsilon +ik\,'+ik}{2}\right),
\end{equation}
which we shall use in Eq. (\ref{U(k,k')}).

\setcounter{equation}{0}
\def\theequation{D.\arabic{equation}}

\section {Continuation of integral operators} 

Since the features of the 
mechanical Liouville model and the free-particle dynamics 
on the half-line $x<0$ are similar, it is sufficient to describe
the procedure of analytical continuation used in section 3.2
for the free-particle case only.
With the required boundary condition $\psi(0)=0$ the Hamilton operator   
$\hat H=-\hbar^2\partial_{xx}^2$ has eigenstates
$\psi_k(x)=\sqrt{{2}/{\pi}}\,\,\sin kx$,
with eigenvalues $\hbar^2k^2$. 
This spectrum is non-degenerated like in Liouville particle dynamics
and we choose $k>0$.

The calculation of matrix elements of exponentials of 
the coordinate operator
 \begin{equation}\label{V_beta}
V_{\beta}(k,k\,')=\langle \psi_k |e^{\beta \hat x}|\psi_{k\,'}\rangle
=\frac{2}{\pi}\int^0_{-\infty} dx\,e^{\beta x}\,\sin kx\, \sin k\,'x
\end{equation}
so simplifies as compared with Eqs. (\ref{e^Q(tau)1}).
The operator $e^{\beta\hat x}$  is bounded for $\beta>0$
and we get
\begin{equation}\label{beta<0}
V_{\beta}(k,k\,')=\frac{1}{\pi}\left(
\frac{\beta}{(k-k\,')^2+\beta^2}-
\frac{\beta}{(k+k\,')^2+\beta^2}\right).
\end{equation}
But for $\beta <0$ the integral (\ref{V_beta}) diverges and
$V_{\beta}(k,k\,')$ has to be considered again as a generalised
function which is a kernel of the operator $e^{\beta\hat x}$ in the
$k$-representation.  So we need in $\beta$ a smooth continuation of
(\ref{beta<0}) acting on test functions $\psi(k)$.  This continuation
does not exist for arbitrary $\psi(k)$. Since $k$ and $ k\,'$ are
positive, the second term of (\ref{beta<0}) is regular for any $\beta$
and has a trivial smooth continuation, whereas a naive continuation of
the first term is singular at $\beta =0$. Indeed, the integral
\begin{equation}\label{I(k)}
I_\beta(\psi\,;\,k)=\frac{1}{\pi}\,
\int_{0}^{\infty}dk\,'\, 
\frac{\beta}{(k\,'-k)^2+\beta^2}\, \, \psi(k\,'),
\end{equation}
has a discontinuity at $\beta =0$
\begin{equation}\label{delta0}
\lim_{\beta\rightarrow \pm 0}\,
I_\beta(\psi\,;\,k)=\pm\, \psi(k).
\end{equation}
 To construct a smooth continuation of the integral (\ref{I(k)})
we split it  into a sum of two integrals
$I_\beta=I_\beta^{(+)} +I_\beta^{(-)}$,
\begin{equation}\label{Ipm}
I_\beta^{(\pm)}(\psi;\,k)
=\pm \frac{1}{2\pi i}\int_0^\infty dk\,'\,\frac{\psi(k\,')}{k\,'-
(k\pm i\beta)},
\end{equation}
and choose a class of test functions $\psi(k)$
which are holomorphic on the half-plane
 $\mbox{Re}\, k>0$ and vanish as $ \mbox{Re}\, k \rightarrow \infty$.
We introduce the Cauchy integrals
\begin{equation}\label{J(k)}
J^{(\pm)}_\beta(\psi;\,k)=\pm\frac{1}{2\pi i}\,
\oint dz\, \frac{\psi(z)}{z-(k \pm i\beta)} ,
\end{equation}
with the integration contours  given in Fig.$\,$4.

\vspace{0.6cm}

\setlength{\unitlength}{.1mm}

\begin{picture}(1500,700)

\put(50,400){\line(1,0){1350}}
\put(60,75){\line(0,1){630}}

\multiput(80,420)(50,0){25}{\line(1,0){25}}
\multiput(80,660)(50,0){25}{\line(1,0){25}}
\multiput(80,420)(0,50){5}{\line(0,1){25}}
\multiput(1315,420)(0,50){5}{\line(0,1){25}}

\multiput(80,380)(50,0){25}{\line(1,0){25}}
\multiput(80,140)(50,0){25}{\line(1,0){25}}
\multiput(80,140)(0,50){5}{\line(0,1){25}}
\multiput(1315,140)(0,50){5}{\line(0,1){25}}

\put(10,400){$0$}
\put(1410,390){$k\,'$}
\put(660,570){$J^+_\beta$}
\put(660,230){$J^-_\beta$}
\put(20,660){$a$}
\put(0,140){$-a$}

\put(700,430){$\longrightarrow$}
\put(700,630){$\longleftarrow$}
\put(90,500){$\downarrow$}
\put(1290,500){$\uparrow$}

\put(700,350){$\longrightarrow$}
\put(700,150){$\longleftarrow$}
\put(90,250){$\uparrow$}
\put(1290,250){$\downarrow$}

\put(920,560){$k\,'=k+i\beta$}
\put(920,260){$k\,'=k-i\beta$}

\put(900,550){$\bullet$}
\put(900,250){$\bullet$}

\put(690,30){$\mbox{Fig.\,4}$}
\end{picture}

\noindent
The Cauchy integrals are well defined for $a>\beta >0$,
where a is a characteristic parameter of the contour.
 In this case  the integrands of (\ref{J(k)}) 
have poles at $z=k \pm i\beta$, and 
$J^{(\pm)}_\beta =\psi(k \pm i\beta)$.
$I_\beta^{(\pm)}$ is then given by
\begin{equation}\label{I+}
I_\beta^{(\pm)}(\psi;\,k)
=\pm\frac{1}{2\pi }\int_0^a d\eta\,\frac{\psi(i\eta)}{-k \pm 
i(\eta-\beta)}
\pm\frac{1}{2\pi i}\int_0^\infty 
d\xi\,\frac{\psi(\xi \pm ia)}{\xi-k\pm i(a-\beta)}
+\psi(k \pm i\beta).
\end{equation}
This form of $I_\beta^{(\pm)}(\psi;\,k)$ obviously has a smooth 
continuation from positive ($0<\beta<a$)
to negative values of $\beta$. Let us denote the continuation
of (\ref{I+}) at 
$\beta =-a$ by $\tilde I^{(\pm)}_{-a}(\psi;\,k)$. 
Since at $\beta =-a$ the contour integrals (\ref{J(k)}) 
vanish, the first two terms  of $\tilde I_{-a}^{(\pm)}$
coincide with $I_{-a}^{(\pm)}$ and one gets
\begin{equation}\label{I+tilde}
\tilde I_{-a}^{(\pm)}(\psi;\,k) =I_{-a}^{(\pm)}(\psi;\,k) +\psi(k\mp ia).
\end{equation}
Thus we have the following smooth set of integral operators 
$\tilde I_\beta$:
\begin{eqnarray}\label{I_epsilon}
\tilde I_{\beta} (\psi;\,k)= \left\{ \begin{array}{cr}
I_{\beta} (\psi;\,k)~~~~~~~~~~~~~~~~~~~~~~~~~~~~~
~~~~~~~~~~~~\,\mbox{for}~~~~~
\beta >0,\\
\psi(k)~~~~~~~~~~~~~~~~~~~~~~~~~~~~~~~~~~~~~~~~~~~~\,\,\,\mbox{for}~~~~~
\beta =0, \\
I_{\beta} (\psi;\,k)+\psi(k+i\beta)+\psi(k-i\beta)
~~~~~~~~~\mbox{for}~~~~~ 
\beta <0. \end{array} \right.
\end{eqnarray}
The kernel (\ref{beta<0}) for negative $\beta$ therefore becomes 
\begin{eqnarray}\label{beta>0}
V_{\beta}(k,k\,')=\frac{1}{\pi}\left(
\frac{\beta}{(k-k\,')^2+\beta^2}-
\frac{\beta}{(k+k\,')^2+\beta^2}\right) \nonumber \\
+ \delta(k-k\,'+i\beta)+\delta(k-k\,'-i\beta).
\end{eqnarray}
This defines the smooth continuation which is applied 
for the calculation of the matrix elements of the vertex operators
of the Liouville particle model in section 3.2.

\section {About hermiticity of $q$-exponentials} 

\setcounter{equation}{0}
\def\theequation{E.\arabic{equation}}

Observables which are not generators of
symmetry transformations can be at most
Hermitean  \cite{RS}.  This holds in particular for the
coordinate operator $i\hbar\partial_p$ on $L^2(\rr_+)$, and
we ask whether the operator $e^{-\gamma\hat q}$ (or
$e^{\gamma\hat q}$) and therefore the vertex (\ref{e^Q=e^q})
can be Hermitean.  

Eq. (\ref{E}) requires for the wave function $\psi(p)\in
L^2(\rr_+)$ an analytical continuation on the complex half-plane
$z=p+i\xi$.  Assuming $\psi(z)\rightarrow 0$  as
$p\rightarrow\infty$, the set of wave functions $\psi(p)$ , ${\cal D}_0$, is dense in 
$L^2(\rr_+)$. But the operator $e^{-\gamma\hat q}$ might be Hermitean 
on a subset ${\cal D}\in{\cal D}_0$ only, and for
any $\psi_1(p),\, \psi_2(p) \in{\cal D}$ hermiticity requires
\begin{equation}\label{herm}
\int_0^{+\infty}dp\,\,
\psi^*_2 (p)\,\psi_1(p - i\hbar\gamma)-
\int_0^{+\infty}dp\,\,
\psi_2^* (p+ i\hbar\gamma)\,\psi_1(p)=0.
\end{equation}
If $\psi (p)\in {\cal D}$ then $\psi^*(p)\in{\cal D} $, and $[\psi(p
-i\hbar\gamma)]^*= \psi^* (p+i\hbar\gamma)$. So
$F(z)=\psi^*_2 (z)\,\psi_1(z-i\hbar\gamma)$ is analytical for $p>0 $
and  
$\oint dz\,F(z)=0$ over the contour of
Fig.{\,}3.

\vspace{6mm}
\setlength{\unitlength}{.1mm}

\begin{picture}(1350,350)

\put(50,60){\line(1,0){1200}}
\put(100,40){\line(0,1){270}}

\multiput(120,80)(50,0){22}{\line(1,0){25}}
\multiput(120,255)(50,0){22}{\line(1,0){25}}
\multiput(120,80)(0,50){4}{\line(0,1){25}}
\multiput(1215,80)(0,50){4}{\line(0,1){25}}

\put(66,70){$0$}
\put(1275,60){$p$}
\put(40,240){$\hbar\gamma$}
\put(660,180){$z=p+i\xi$}

\put(700,81){$\longrightarrow$}
\put(700,235){$\longleftarrow$}
\put(120,165){$\downarrow$}
\put(1190,165){$\uparrow$}

\put(132,93){$_1$}
\put(1190,93){$_2$}
\put(1190,225){$_3$}
\put(132,225){$_4$}
\put(140,38){$\epsilon$}
\put(117,50){$\bullet$}

\put(1203,35){$_A$}
\put(1206,50){$\bullet$}

\put(87,240){$\bullet$}

\put(630,10){$\mbox{Fig.\,3}$}
\put(96,330){$\xi$}
\end{picture}

\noindent
For $\epsilon\rightarrow 0$ and
$A\rightarrow\infty$ the contributions of (1,2) and (3,4) 
to the contour integral
coincide with the first respectively second term of 
(\ref{herm}).  Since the line $(2,3)$ 
does not contribute, $\oint dz\,F(z)=0$ gives 
 for a Hermitean action of $e^{-\gamma\hat q}$
on the Hilbert space the condition
 \begin{equation}\label{herm1}
\lim _{\epsilon \rightarrow 0}\int_0^{\hbar\gamma} d\xi\,\,
\psi_2^* (i\xi+\epsilon)\,\psi_1(i\xi -i\hbar\gamma+\epsilon)=0.
\end{equation}
This requires for $\psi(z)$ a definite boundary behaviour on the half-plane.
In case we would assume regularity at the boundary $p=0$,
$\lim_{\epsilon \rightarrow 0} \psi(\epsilon -i\xi)=f(\xi)$,
the hermiticity condition (\ref{herm1}) becomes
 \begin{equation}\label{herm2}
\int_0^{\hbar\gamma} d\xi\,\,
f_2^* (\xi)\,f_1(\hbar\gamma-\xi)=0.
\end{equation}
This hermiticity condition is obviously satisfied by the periodic
functions
\begin{equation}\label{functions}
f(\xi)=\sum_{n>0} c_n \,
e^{-i\frac{2\pi n}{\hbar\gamma}\xi},
\end{equation}
but they are not invariant under time evolution $\psi(p)\mapsto
e^{-i\frac{p^2\tau}{4\pi\hbar}}\,\,\psi(p)$, which requires
 \begin{equation}\label{herm3}
\int_0^{\hbar\gamma} d\xi\,\,
f_2^* (\xi)\,f_1(\hbar\gamma-\xi)e^{i\frac{\gamma\xi}{2\pi}\tau}  =0.
\end{equation}
For $f_1(\xi)=f_2(\xi)=f(\xi)$, then $f(\xi)=0$,
at least, on a half of the interval $\xi\in (0, \hbar\gamma)$.

From this result one easily obtains properties for the wave
functions $\psi(p)$, if we transform 
by
\begin{equation}\label{map}
\zeta =\frac{z-1}{z+1}
\end{equation}
the half-plane $\mbox{Re}\, z>0$ onto the interior 
of the unit disk $|\zeta|<1$,
the boundary $\mbox{Re}\, z=0$ to the unit circle 
$|\zeta|=1$, and exclude the point $\zeta =1$.
A theorem \cite{Tit} tells us that if an analytical function $f(\zeta)$ is 
bounded for $|\zeta|<1$, and $f(\zeta)\rightarrow 0$ as 
$|\zeta|\rightarrow 1$ on a continuous part of $|\zeta|=1$,
then $f(\zeta)$ is identically zero, i.e., $\psi (p)=0$.

We conclude that the hermiticity condition (\ref{herm1})
has a solution only if 
the analytical functions $\psi(z)$ do not have a regular limit at the
boundary $\mbox{Re}\, z=0$, and cannot be continued either out
of the half-plane.  Such analytical functions are called
`functions with natural boundaries' \cite{Tit}, but so far we do not
have the solution for such a function $\psi(p)$ .

\end{appendix}

\end{document}